\documentclass[preprint,12pt]{elsarticle}

\usepackage{amssymb}
\usepackage{amsmath}
\usepackage{enumitem} 
\usepackage{zhlipsum} 
\usepackage{amssymb,amsmath,amsthm,amsfonts}
\usepackage{verbatim}
\usepackage{mathrsfs}
\usepackage{amsmath}
\usepackage[colorlinks=true,linkcolor=red,citecolor=red,urlcolor=blue,]{hyperref}
\usepackage[pagewise]{lineno}\nolinenumbers 
\usepackage{cleveref}
\usepackage{graphicx}
\usepackage{epstopdf}
\usepackage{adjustbox}
\usepackage{longtable}
\usepackage{caption}
\usepackage{float}
\usepackage{underscore}

\journal{Neurocomputing}

\begin{document}

\begin{frontmatter}



\title{When Mamba Meets xLSTM: An Efficient and Precise Method with the xLSTM-VMUNet Model for Skin lesion Segmentation}

\author[label1]{Zhuoyi Fang, Jiajia Liu, Kexuan Shi, Qiang Han$^{\ast}$}
\affiliation[label1]{organization={School of Mathematical Sciences},
           addressline={Yangzhou University},
          city={Yangzhou},
          postcode={225009},
          country={China,\quad},
          email={Email: hanqiang@yzu.edu.cn}}
\begin{abstract}
Automatic melanoma segmentation is essential for early skin cancer detection, yet challenges arise from the heterogeneity of melanoma, as well as interfering factors like blurred boundaries, low contrast, and imaging artifacts. While numerous algorithms have been developed to address these issues, previous approaches have often overlooked the need to jointly capture spatial and sequential features within dermatological images. This limitation hampers segmentation accuracy, especially in cases with indistinct borders or structurally similar lesions. Additionally, previous models lacked both a global receptive field and high computational efficiency. In this work, we present the xLSTM-VMUNet Model, which jointly capture spatial and sequential features within dermatological images successfully. xLSTM-VMUNet can not only specialize in extracting spatial features from images, focusing on the structural characteristics of skin lesions, but also enhance contextual understanding, allowing more effective handling of complex medical image structures. Experiment results on the ISIC2018 dataset demonstrate that xLSTM-VMUNet outperforms VMUNet by 4.00\% on DSC and 6.93\% on IoU on the ISIC2017 dataset, by 1.25\% on DSC and 2.07\% on IoU on the ISIC2018 dataset, with faster convergence and consistently high segmentation performance.
\end{abstract}


\end{frontmatter}



\section{Introduction}
As the largest organ in the human body, the skin acts as the primary barrier against ultraviolet radiation, safeguarding the body from its detrimental effects \cite{1}. The 2023 global cancer statistics indicate that malignant skin lesions account for tens of thousands of deaths each year \cite{2}. Notably, melanoma, a highly aggressive type of skin cancer, is rapidly becoming one of the fastest-increasing cancers around the world. Segmentation plays a vital and challenging role in the workflow for automated skin lesion analysis. In recent years, a large number of computer-assisted segmentation techniques have been developed for medical images. However, segmenting medical image automatically and accuratelly is a literally challenging task because these images are inherently complex and rarely contain simple linear features. Over the past decade, numerous studies have focused on the development of efficient and robust segmentation methods for medical imaging. Among these contributions, U-Net stands out as a seminal work that first illustrated the efficacy of encoder-decoder convolutional networks with skip connections for medical image segmentation, and it has also yielded promising outcomes in various image translation tasks. Since the inception of U-Net \cite{16}, a number of significant modifications have emerged \cite{13,14,15}, particularly within the domain of medical imaging, including variants such as U-Net++ \cite{17}, 3D U-Net \cite{18}, V-Net \cite{19}, and Ynet \cite{20}. Additionally, UneXt \cite{21}, Rolling-UNet \cite{22}, HEA-Net \cite{23} and U-MLP \cite{24} incorporate hybrid methodologies that combine convolutional operations with multi-layer perceptrons (MLP) to enhance the performance of segmentation networks, thereby facilitating their application in resource-limited point-of-care settings. Recently, a variety of networks leveraging Convolutional Neural Networks (CNN) and Vision Transformers (ViT) have been employed to augment the U-Net architecture for medical image segmentation. These networks have proven effective in addressing global context and long-range dependencies within the segmentation tasks. In comparison to CNN, ViT typically exhibits enhanced learning capabilities on large-scale medical image segmentation datasets, attributed to its incorporation of the self-attention mechanism. However, the quadratic complexity associated with the self-attention mechanism, coupled with the substantial number of tokens, results in significant computational overhead when applied to large-scale skin lesion image segmentation tasks that involve high spatial resolutions. The unresolved trade-off between achieving a global receptive field and maintaining high computational efficiency drives the need for a novel architecture tailored for large-scale skin lesion image segmentation, which aims to retain the intrinsic benefits of the standard self-attention mechanism, including global receptive fields and dynamic weighting parameters.\\
\indent{State Space Models (SSM) \cite{39,40,41,42} have attracted significant interest owing to their computational efficiency Mamba \cite{43}, a novel SSM within the realm of natural language processing (NLP) \cite{44,45}, has emerged as a highly promising method for modeling long sequences with linear complexity \cite{46}. Due to its advantages over conventional foundational architectures, Mamba holds significant potential as a visual foundational architecture \cite{47,48}. It has been actively utilized across various computer vision tasks \cite{49}, contributing notably to the field of medical image segmentation \cite{50}. Inspired by Mamba, Vision Mamba \cite{51} and Visual Mamba \cite{52} were the pioneering implementations of this model in the visual domain, achieving very impressive results. Furthermore, several researchers have adapted it for medical image segmentation, resulting in innovations such as VM-Unet \cite{53}, U-Mamba \cite{54}, H-vmunet \cite{55}, Swin-Umamba \cite{56}, nnMamba \cite{57}, Mamba-UNet \cite{58}, LightM-Unet \cite{59} and UltraLight-VM-UNet \cite{60}. Particularly, SkinMamba \cite{61} facilitates expert knowledge exchange across different levels in a global state and achieves high-frequency restoration and boundary prior guidance.}\\
\indent{However, the challenges remaining in skin lesion segmentation have not been solved completely. For instance, existing models exhibit limitations in simultaneously capturing spatial and sequential features, which hampers their ability to comprehensively understand the information within skin lesion images. Additionally, these models lack contextual understanding, making it challenging to establish long-range dependencies when processing skin lesion images. Previous methods also struggle to achieve precise segmentation in the presence of ambiguous boundaries or similar structures, resulting in inadequate extraction of detailed spatial features, particularly in addressing subtle structures and lesion boundaries. Furthermore, these methods find it difficult to balance global receptive fields with high computational efficiency.}\\
\indent{Recently, the Extended Long Short-Term Memory (xLSTM) \cite{62} architecture represents a significant advancement in language modeling, demonstrating potential to compete with Transformer and State SSM. By optimizing the performance of Long Short-Term Memory (LSTM), xLSTM effectively manages long-range dependencies while maintaining linear computational and memory efficiency. Similar to models such as ViT in the field of computer vision, xLSTM offers a powerful alternative in the evolving landscape of medical image segmentation. Vision-LSTM \cite{73} employed a generic computer vision backbone that uses xLSTM blocks as its core components. xLSTM-UNet \cite{74} proposed the xLSTM-enabled U-Net image segmentation network that can perform both 2D and 3D medical image segmentation tasks and achieves state-of-the-art (SOTA) results.}\\
\indent{In this paper, we propose the xLSTM-VMUNet model, which combines Mamba with xLSTM for efficient skin lesion segmentation. Through a carefully designed architecture, our model excels in feature extraction, memory retention, adaptability, and nonlinear modeling. First, we leverage Mamba for deep spatial feature extraction using a multi-layer convolutional and pooling network to progressively capture essential features from skin lesion images. The integration of residual connections enhances feature transfer efficiency and improves processing capability for complex lesion patterns, establishing a robust basis for xLSTM. Following feature extraction, we input spatial features into xLSTM to utilize its long-term and short-term memory mechanisms. xLSTM processes sequential data while preserving crucial contextual information via a detailed state update mechanism. Specifically, we designed gating mechanisms that allow the model to selectively retain key features during skin lesion processing, maintaining segmentation performance across varying lesion types. To boost adaptability, we incorporated structured input processing in xLSTM, including a BlockDiagonal structure, to address diverse skin lesion types. This dynamic feature adjustment enables the model to adapt flexibly to the varying characteristics and distributions of different lesion categories, a crucial capability given the diverse morphology and sizes found in skin lesions. Finally, for feature fusion, we implemented a multi-level feature integration mechanism, combining the extracted features by Mamba with the output of xLSTM through concatenation and weighted fusion. This strategy enhances segmentation accuracy and improves pattern recognition. We employed cross-validation during validation to ensure stability and reliability across different skin lesion datasets.}\\
\indent{In summary, the following are the major contributions of our work:}
\begin{enumerate}[label=\arabic*)] 
	\item We introduce the xLSTM-VMUNet model, which is designed to efficiently extract and integrate both spatial features and temporal information within skin lesion images. By leveraging the strengths of xLSTM in modeling long-range dependencies and temporal dynamics, our xLSTM-VMUNet model significantly improves the accuracy and robustness of the segmentation process. 
	\item We propose a multi-level feature integration mechanism to effectively combine the features extracted by Mamba with the processing results of xLSTM. The feature linking and weighted fusion ensures that feature information at different levels can be fully integrated, which significantly improves the accuracy of the model for skin lesion segmentation and enhances the ability of the model to capture complex patterns.
	\item We conduct extensive experiments conducted on the ISIC2018 skin lesion segmentation dataset, demonstrating that xLSTM-VMUNet outperforms. SOTA (state-of-the-art) methods across Dice and IoU metrics, with faster convergence and consistent high segmentation performance.
\end{enumerate}
\section{Related Work}

\subsection{Medical image segmentation}
Medical images play a pivotal role in aiding healthcare providers in diagnosis and treatment decisions. Interpretation relies on radiologists' expertise, which is time-intensive and subjective, influenced by experience and training. Therefore, integrating computer-aided systems has become crucial to ensure efficient, objective, and consistent analysis of medical images. Image segmentation, a key process in medical physics, is essential for identifying tumors or lesions by partitioning an image into homogeneous regions for extracting diagnostic information. Advanced techniques have been developed to address the limitations of traditional methods, significantly improving both accuracy and efficiency in medical image analysis.\\
\indent{Since the introduction of U-Net, several variants have been proposed to enhance its performance. U-Net++ replaces traditional cropping and concatenation with dense convolutions, improving feature fusion and reducing information loss. Attention-U-Net integrates attention mechanisms to focus on relevant regions and suppress irrelevant information. Res-U-Net introduces residual blocks to stabilize feature transfer and improve deep network training. Dense-U-Net employs Dense-blocks for efficient feature reuse and better multi-scale information capture. U-Netv2 uses an innovative skip connection mechanism to refine feature fusion and improve integration across scales.}

\subsection{Mamba}
Mamba, based on the SSM, provides a more efficient alternative to transformers in medical image analysis. With linear time complexity, it processes longer sequences faster, reduces memory usage, and excels in multimodal data integration, enhancing diagnostic accuracy and patient outcomes.\\ 
\indent{Vim (Zhu et al., 2024a) is a Mamba-based architecture that processes image patch sequences, using position embeddings and a class token. The sequence is then passed through Vim blocks with both forward and backward SSM paths. VMamba (Liu et al., 2024g) enhances Mamba’s 1D scanning by introducing the Cross-Scan Module, which scans image patches in four directions and applies selective SSMs to capture cross-directional dependencies. The sequences are merged back into the original 2D layout. It uses stacked VSS blocks with down-sampling, where the vanilla VSS block replaces 1D convolution with 2D depthwise convolution and incorporates SS2D with layer normalization. PlainMamba (Yang et al., 2024a) is a non-hierarchical architecture optimized for multi-scale fusion, multi-modal integration, and hardware efficiency. It processes 2D patches with depthwise convolutions and adapted selective scanning using zigzag and direction-aware updates.}\\
\indent{LocalMamba (Huang et al., 2024e) resolves local token dependency issues by dividing the image into local windows for directional selective scanning (SSM), while maintaining global SSM. It also uses spatial and channel attention before patch merging and optimizes scan directions for each layer to improve efficiency. Efficient-VMamba (Pei et al., 2024) combines Efficient 2D Scanning (ES2D) with atrous sampling and a convolutional branch for global and local feature extraction, processed by a Squeeze-and-Excitation (SE) block. The output forms the EVSS block, with EVSS in early stages and Inverted Residual blocks in later stages.}
\subsection{xLSTM}
The xLSTM model marks a significant breakthrough in the field of sequence modeling, offering a powerful alternative to established architectures like Transformer and SSM, which have dominated machine learning applications in natural language processing and computer vision. Building upon traditional Long Short-Term Memory (LSTM) networks, xLSTM introduces key innovations that enable efficient capture of long-range dependencies, all while maintaining linear computational complexity and memory efficiency. This makes xLSTM particularly well-suited for scenarios involving large-scale datasets and high-dimensional inputs, where conventional RNN models often encounter scalability and performance limitations.\\ 
\indent{xLSTM presents a promising alternative to traditional convolutional methods in medical image segmentation, a field that demands high precision and effective multi-scale feature integration. The Vision-LSTM model exemplifies this approach by embedding xLSTM blocks into a general-purpose computer vision architecture, improving the model's ability to capture both spatial and temporal dependencies within medical images. This integration enhances feature representation, enabling more accurate differentiation of subtle patterns, such as tumor structures and organ boundaries.}\\
\indent{Furthermore, the xLSTM-UNet architecture, which integrates xLSTM units with the U-Net framework, has demonstrated outstanding performance in both 2D and 3D medical image segmentation tasks. By harnessing xLSTM's ability to model long-range dependencies, xLSTM-UNet enhances the network's capacity to capture global contextual information, thereby improving segmentation accuracy, particularly in complex cases with intricate anatomical structures or heterogeneous tissue types.}
\begin{figure}[h]
	\centering
	\includegraphics[width=1.1\linewidth]{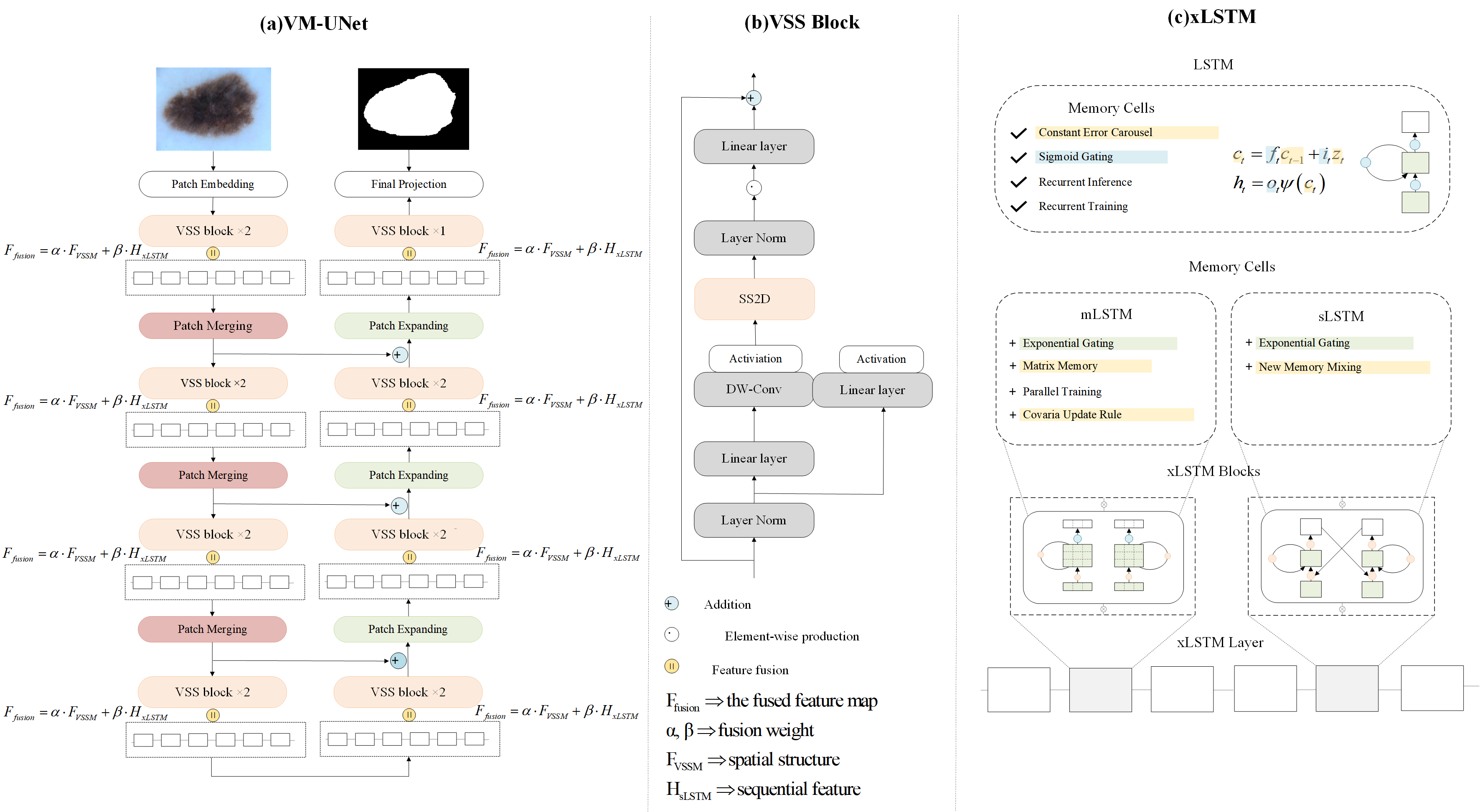}
	\caption{The Architecture Overview of XLSTM-VMUNet.}
\end{figure}
\section{Method}
xLSTM_VMUNet is a deep learning framework that integrates the Visual State Space Model (VSSM) with an extended Long Short-Term Memory network (xLSTM) to enhance the accuracy and robustness of medical image segmentation. As the primary feature extraction module, VSSM is responsible for constructing spatial representations through a hierarchical feature extraction mechanism that captures multi-scale information from the input image in a bottom-up manner. Its encoder progressively downsamples the input to construct high-level semantic features while leveraging the State Space Model (SSM) to optimize information propagation, thereby combining the local perceptual capabilities of convolutional networks with the long-range dependency modeling ability of transformer architectures. During the decoding phase, VSSM restores the spatial resolution of the extracted features through upsampling and integrates shallow and deep semantic information using skip connections, ensuring precise boundary reconstruction of target regions. However, despite its effectiveness in spatial feature extraction, VSSM exhibits certain limitations in capturing the complex interactions between local patterns and global dependencies. This is particularly crucial in medical image segmentation, where strong sequential modeling capabilities are required to analyze relationships across different scales and anatomical structures. To address this, xLSTM_VMUNet introduces xLSTM as a post-processing module to enhance feature consistency and global representation, building upon the high-dimensional feature maps extracted by VSSM.
\\
\hspace*{1em}
As the core sequence modeling unit of xLSTM_VMUNet, xLSTM integrates two distinct long short-term memory structures, namely structured LSTM (sLSTM) and multi-scale LSTM (mLSTM), to meet the demands of hierarchical feature modeling. sLSTM employs a block-diagonal weight matrix to achieve efficient gating mechanisms while incorporating causal convolution to enhance the extraction of local patterns, enabling effective transmission of key features over short-range dependencies. In contrast, mLSTM employs a multi-scale attention mechanism that dynamically weights features across different time steps, thereby establishing global dependencies over longer time spans. These two LSTM variants work in synergy, with sLSTM focusing on fine-grained local information modeling and mLSTM reinforcing feature coherence through multi-scale interactions, effectively compensating for the shortcomings of VSSM in feature representation. Through this dual modeling strategy, xLSTM allows features to adaptively capture complex structural patterns in high-dimensional space, improving segmentation performance by enhancing the delineation of pathological regions and anatomical boundaries.
\\
\hspace*{1em}
Within the overall xLSTM_VMUNet architecture, the interaction between VSSM and xLSTM is a critical component that ensures the model’s performance. VSSM, as the primary feature extractor, converts the input image into multi-scale feature maps and restores spatial information through skip connections in the decoding phase. These high-dimensional features are then rearranged to accommodate the sequential modeling requirements of xLSTM, transforming the two-dimensional feature maps into one-dimensional sequences that enable xLSTM to process and enhance different-scale features through its gating units. This transformation allows the generated static features by VSSM to be further refined in the temporal dimension, ensuring that the model not only learns local patterns but also comprehensively understands the dynamic evolution of global structures. Moreover, VSSM provides hierarchical information through its layer-wise feature extraction, while xLSTM dynamically updates these features through its state transition mechanisms, ensuring their coherence in high-dimensional space. This integration not only preserves spatial feature integrity but also facilitates efficient fusion of different hierarchical representations, providing stable and precise inputs for the subsequent segmentation task.
\\
\hspace*{1em}
The advantages of xLSTM_VMUNet in skin lesion Segmentation are primarily reflected in its ability to analyze complex pathological patterns. Traditional U-Net and its variants typically rely on convolutional operations for feature extraction, but their limited receptive field constrains their ability to capture long-range dependencies, whereas transformer-based architectures often require extensive data for training and incur high computational costs. In contrast, xLSTM_VMUNet achieves efficient modeling of both short-range and long-range dependencies by leveraging the multi-scale spatial features provided by VSSM and the sequential modeling capabilities of xLSTM, enabling the extraction of critical information from pathological regions at various scales, thereby improving segmentation accuracy. Furthermore, by combining sLSTM and mLSTM, the model adaptively integrates information across different time scales, enhancing its ability to delineate ambiguous boundaries and complex morphologies, particularly in skin lesion images with low contrast or significant textural variations. In summary, xLSTM_VMUNet successfully constructs a segmentation framework that combines efficient feature extraction with robust sequence modeling through the innovative integration of VSSM and xLSTM, providing a novel solution for medical image analysis.
\\
\hspace*{1em}
We elaborate on the model's preparatory theory and model details. Firstly, we introduce selective state space models (S6). Then, we describe the VSS block and the xLSTM block. Finally, we ascensively elaborate on the core structure: the xLSTM-VMUNet Model.

\subsection{SSM}
SSM is a core class of framework in deep learning for processing sequential data. These models aim to map an input sequence, denoted as $x(t)$, from a real vector space $R^L$ to an output sequence $y(t)$ within the same space, while leveraging an intermediate latent state $h(t)$ also within $R^N$. The system dynamics are governed b a series of linear transformations, as outlined in the corresponding set of equations:
\begin{align}\label{1}
  \left\{
\begin{array}{lll}
	dh(t)/dt = Ah(t)+Bx(t),& \\
	y(t) = Ch(t),&
\end{array}
\right.
\end{align}
where $A$, $B$ and $C$ are system matrices of suitable dimensions that govern the state transition dynamics, the influence of the input, and the mapping to the output, respectively. These matrices are defined as $A\in\mathbb{R}^{N\times{N}}$, $B\in\mathbb{R}^{N\times{1}}$, and $C\in\mathbb{R}^{N\times{1}}$.\\
\indent{For practical implementation, the continuous-time model is discretized using a zero-order hold approximation, converting the system matrices $A$ and $B$ into their discrete counterparts over a sampling period $\Delta$. The discretized system is represented as:}
\begin{align}\label{2}
	\left\{
	\begin{array}{ll}
	\overline{A} = exp(\Delta{A}),& \\
		\overline{B} = \Delta{A}^{-1}(exp(\Delta{A})-I)\cdot\Delta{B}.&
	\end{array}
	\right.
\end{align}
\indent{The resulting equations for the discrete model are given by:}
\begin{align}\label{3}
	\left\{
	\begin{array}{ll}
		h_t = \overline{A}h_{t-1}+\overline{B}x_t,& \\
		y_t = Ch_{t},&
	\end{array}
	\right.
\end{align}
\indent{To improve efficiency, the entire sequence output can be computed simultaneously using a global convolution, boosting both scalability and speed. This is formulated as:}
\begin{align}\label{4}
	\left\{
	\begin{array}{ll}
		y = x\otimes\overline{K},& \\
		\overline{k} = (C\overline{B},C\overline{AB},...,C\overline{A}^{L-1}\overline{B}),&
	\end{array}
	\right.
\end{align}
where $\otimes$ denotes the convolution operation, $L$ represents the sequence length, and $\overline{K}$ is the kernel derived from the SSM, specifically tailored for efficient sequence processing.

\subsection{Selective State Space Models (S6)}
The linear time-invariant state-space model treats all tokens equally, neglecting dynamic content importance. Prioritizing more relevant tokens and adjusting attention accordingly is more effective for complex inputs.\\
\indent{Mamba integrates a selectivity mechanism into the state-space model, creating Selective State Space Models (S6). \\
\indent{In Mamba, the matrices $B$, $C$ and $\Delta$ are input-dependent, allowing adaptive behavior. The discretization process with the selectivity mechanism is as follows:}
\begin{align}\label{5}
	\left\{
	\begin{array}{lll}
		\overline{B} = s_{B}(x),& \\
		\overline{C} = s_{C}(x),& \\
		\Delta = \tau_{A}(\Delta+s_{A}(x)),&
	\end{array}
	\right.
\end{align}
where $\overline{B}\in\mathbb{R}^{B\times{L}\times{N}}$, $\overline{C}\in\mathbb{R}^{B\times{L}\times{N}}$ and $\overline{\Delta}\in\mathbb{R}^{B\times{L}\times{D}}$. $s_{B}(x)$ and $s_{C}(x)$ are linear functions that project the input $x$ into a N-dimensional space, while $s_{A}(x)$ projects the hidden state dimension $D$ linearly into the desired dimension, connected to the RNN gating mechanism. These computations transform $\Delta$, $B$ and $C$ into input-dependent functions of length $L$, converting the time-invariant model into a time-varying one and enabling selectivity.\\
\indent{The parameter $\Delta$ is expanded to $(B,L,D)$, giving each token in a batch a unique input-dependent control. A larger step size for $\Delta$ prioritizes the input, while a smaller one emphasizes the stored state. Parameters $B$ and $C$ become input-dependent, refining the control between input $x$, state $h$, and output $y$. While $A$ remains independent, its relevance to the input is introduced via data dependency of $\Delta$. With dimension $N$, $A$ adapts across SSM dimensions, enabling precise generalization.}
\newcommand{\Myref}[1]{Fig.\ref{#1}}
\subsection{VSS Block}
The VSS Block, derived from VMamba, is central to XLSTM-VMUNet. After Layer Normalization, the input splits into two branches: one applies a linear layer, the other adds depthwise separable convolution and uses the 2D-Selective-Scan (SS2D) module. The branches are fused, normalized, and merged with a residual connection, using SiLU as the activation.\\
\indent{SS2D unfolds the input in four directions, processes it in the S6 block, and merges it back, enhancing selectivity by adjusting parameters to capture key features and reduce noise. \Myref{Fig.2} reveals the scan expanding operation and the scan merging operation in SS2D.}
\begin{figure}[h]
	\centering
	\includegraphics[width=1.1\linewidth]{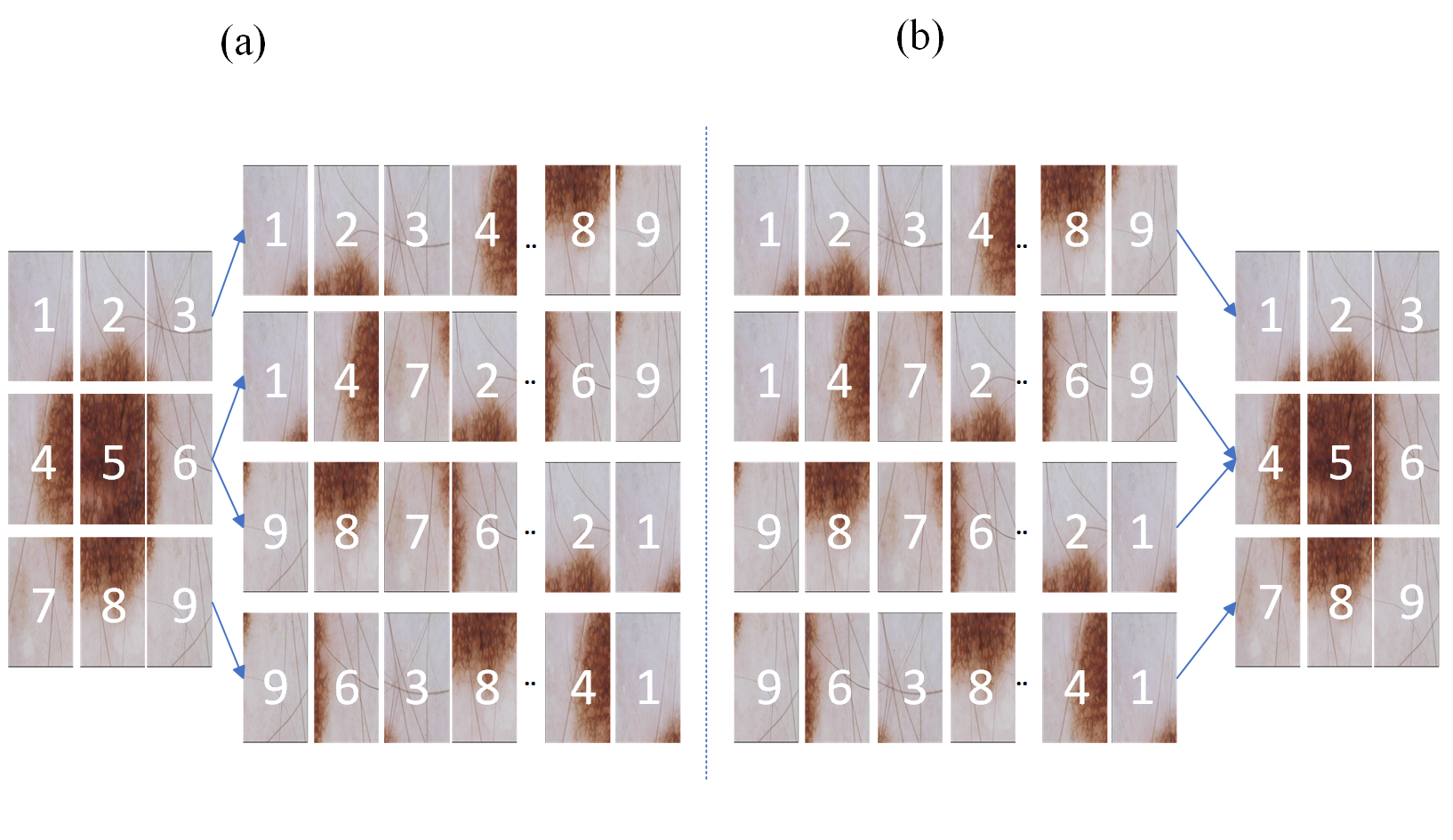}
	\caption{(a) The scan expanding operation in SS2D. (b) The scan merging operation in SS2D.}\label{Fig.2}
\end{figure}
\subsection{xLSTM Block}
\subsubsection{sLSTM Block}
The sLSTM Block extends the traditional LSTM architecture by introducing exponential gating and norma-lization states to enhance the control over information storage and flow.\\
\indent{Firstly, we introduce the memory cell module and state update process. The memory cell is updated as follows:}
\begin{align}\label{6}
	\left\{
	\begin{array}{ll}
		c_t = f_{t}c_{t-1}+i_{t}z_{t},& \\
		z_t = tanh(W_{z}x_{t}+R_{z}h_{t-1}),&
	\end{array}
	\right.
\end{align}
where $c_t$ is the memory cell at time step $t$. $f_t$ and $i_t$ are the forget and input gates respectively. $z_t$ is the candidate memory, controlled by $W_z$ and $R_z$.\\
\indent{Normalization state update:}\\
\begin{align}\label{7}
    n_t=f_{t}n_{t-1}+i_{t},
\end{align}
where $n_t$ is the normalization state that balances the contribution of the forget and input gates.\\
\indent{Then, we introduce the projection and residual connection. The hidden state is computed as:}
\begin{align}\label{8}
	\left\{
	\begin{array}{ll}
		h_t = o_{t}c_{t}/n_{t},& \\
		o_t = \sigma(W_{o}x_{t}+R_{o}h_{t}),&
	\end{array}
	\right.
\end{align}
where $o_{t}$ is the output gate, controlling the final output $h_t$. Normalizing $c_t$ by $n_t$ ensures numerical stability.\\
\indent{The hidden state is further processed through up-projection, non-linear transformation, and down-projection:}
\begin{align}\label{9}
	\left\{
	\begin{array}{llll}
		y_{left} = W_{up-left}h_t,& \\
		y_{right} = W_{up-right}h_t,&\\
		y_{gated} = GELU(y_{right}),&\\
		y_{out} = W_{down}(y_{left}\cdot y_{gated}).&
	\end{array}
	\right.
\end{align}
\indent{The final output includes a residual connection:}
\begin{align}\label{10}
	F=y_{out}+x,
\end{align}
where $F$ represents the final output.

\subsubsection{mLSTM Block}
The mLSTM Block augments memory capacity by extending the memory cell from a scalar to a matrix, thereby enabling more complex storage and representation.\\
\indent{The memory cell is updated in matrix form:}
	\begin{align}\label{11}
		\left\{
		\begin{array}{lll}
			C_t = f_t C_{t - 1}+i_t v_t k_t^{\top},&\\
			k_t = \frac{1}{\sqrt{d}}W_k x_t + b_k,&\\
			v_t = W_v x_t + b_v,&
		\end{array}
		\right.
	\end{align}
where $C_t$ is the memory matrix, $v_t$ and $k_t$ are the value and key vectors respectively. $W_k$, $W_v$ are the weights for generating key and value vectors. $b_k$, $b_v$ are the biases.\\
\indent{To retrieve information from the matrix memory, mLSTMBlock uses a query vector $q_t$:
\begin{align}\label{12}
	\left\{
	\begin{array}{lll}
		h_t = o_t \odot \frac{C_t q_t}{\max(|n_t^{\top} q_t|, 1)},&\\
		q_t = W_q x_t + b_q,&\\
		n_t = f_t n_{t - 1}+i_t k_t,&
	\end{array}
	\right.
\end{align}
where $q_t$ is the query vector and $n_t$ is the normalization state.\\
\indent{The matrix-based memory update and the retrieval mechanism significantly enhance the ability of the model to capture complex temporal relationships. By effectively managing and retaining information across time steps, it preserves long-term dependencies. This dynamic memory adjustment improves the capacity of the model to model intricate temporal patterns, which is crucial for tasks involving sequential data, where understanding complex dependencies is essential for accurate modeling.}
\subsection{xLSTM-VMUNet}
\subsubsection{VSSM Feature Extraction}
The architecture integrates the VSSM with the xLSTM component through the xLSTM-VMUNet framework. The Visual Structured State Model (VSSM) is responsible for extracting deep spatial features from the input image. Given an input image $x\in\mathbb{R}^{H\times W\times C}$, VSSM utilizes multiple convolution and pooling layers to progressively extract essential features.\\
\indent{For a given layer $l$, let $F^{(l)}$ denote the feature map after convolution. The convolution operation in VSSM is defined as:}
\begin{equation}\label{13}
	F^{(l)}=\sigma\left(W^{(l)}\otimes F^{(l - 1)}+b^{(l)}\right),
\end{equation}
where $W^{(l)}$ represents the weight matrix of layer $l$, $b^{(l)}$ is the bias term for layer $l$, $\otimes$ denotes the convolution operation, $\sigma$ is the activation function (e.g., ReLU), and $x$ is the input image.\\
\indent{Through these layered convolutions, VSSM produces a set of multi-scale spatial features, denoted as $\{F_{VSSM}^{(1)}, F_{VSSM}^{(2)},..., F_{VSSM}^{(L)}\}$, where each $F_{VSSM}^{(l)}$ spatial information at different scales, providing rich representations for subsequent sequence modeling.}
\subsubsection{Sequence Modeling with xLSTM}
Following spatial feature extraction, the xLSTM network is employed to model long short-term dependencies. Let the spatial features from VSSM be $F_{VSSM}=\{ F_{VSSM}^{(1)}, F_{VSSM}^{(2)},..., F_{VSSM}^{(L)} \}$, where each $F_{VSSM}^{(l)}$ represents a timestep in the sequence.\\
\indent{The xLSTM state update equations are:}
\begin{align}\label{14}
	\left\{
	\begin{array}{llll}
i_t = \sigma\left(W_i \cdot F_{VSSM}^{(t)}+U_i \cdot h_{t - 1}+b_i\right),&\\
f_t = \sigma\left(W_f \cdot F_{VSSM}^{(t)}+U_f \cdot h_{t - 1}+b_f\right),&\\
o_t = \sigma\left(W_o \cdot F_{VSSM}^{(t)}+U_o \cdot h_{t - 1}+b_o\right),&\\
g_t = \tanh\left(W_g \cdot F_{VSSM}^{(t)}+U_g \cdot h_{t - 1}+b_g\right),&
	\end{array}
	\right.
\end{align}
where $i_t$, $f_t$ and $o_t$ denote the input gate, forget gate, and output gate at timestep t respectively, $g_t$ represents the candidate state, $W_i$, $W_f$, $W_o$, $W_g$ are weight matrices for the gates, $U_i$, $U_f$, $U_o$, $U_g$ are weight matrices associated with the hidden state $h_{t-1}$ from the previous timestep, $b_i$, $b_f$, $b_o$, $b_g$ are bias vectors, $\sigma$ is the sigmoid activation function.\\
\indent{The cell state $c_t$ and hidden state $h_t$ are updated as follows:}
\begin{align}\label{15}
	\left\{
	\begin{array}{ll}
		c_t = f_t \odot c_{t - 1}+i_t \odot g_t,&\\
		h_t = o_t \odot \tanh(c_t),&
	\end{array}
	\right.
\end{align}
where $c_t$ is the cell state at timestep $t$, $h_t$ represents the hidden state at timestep $t$, $\odot$ denotes the element-wise multiplication, $tanh$ is the hyperbolic tangent activation function.\\
\indent{This gating mechanism enables xLSTM to selectively retain or forget spatial features, which is critical for preserving long-term dependencies and contextual information across sequential image data.}
\subsubsection{Multi-level Feature Fusion}
The model employs a multi-level feature fusion mechanism to combine the spatial features from VSSM and temporal features from xLSTM, resulting in a comprehensive representation. Let $F_{VSSM}$ be the spatial feature map from VSSM, and $H_{xLSTM}$ be the hidden state output from xLSTM. The fused feature map $F_{fusion}$ is given by:
\begin{equation}\label{16}
	F_{fusion}=\alpha\cdot F_{VSSM}+\beta\cdot H_{xLSTM},
\end{equation}
where $F_{fusion}$ represents the fused feature map that combines spatial and temporal information, $\alpha$ and $\beta$ are fusion weights, determining the contributions of $F_{VSSM}$ and $H_{xLSTM}$ to the fused feature, $F_{VSSM}$ captures spatial structure, and $H_{xLSTM}$ captures sequential and memory-dependent features.\\
\indent{This fusion strategy enables the model to capture complex patterns across both spatial and temporal domains, enhancing segmentation performance on challenging data.}
\newcommand{\myref}[1]{Eq.\ref{#1}}
\subsubsection{Loss function}
The introduction of xLSTM-VMUNet is designed to evaluate the potential benefits of combining Mamba with xLSTM for more efficient skin lesion segmentation. In this context, we focus solely on the most fundamental loss functions, namely Binary Cross-Entropy and Dice loss (denoted as BceDice loss), for the skin lesion segmentation task. These loss functions are formally represented in \myref{eq.17} and \myref{eq.18}.
\begin{equation}\label{eq.17}
	L_{BceDice}=\lambda_{1}L_{Bce}+\lambda_{2}L_{Dice},
\end{equation}
\begin{align}\label{eq.18}
	\left\{
	\begin{array}{ll}
		L_{Bce}=-\frac{1}{N}\sum_{i = 1}^{N}[y_i \log(\hat{y}_i)+(1 - y_i)\log(1 - \hat{y}_i)],\\
		L_{Dice}=1-\frac{2|X\cap Y|}{|X|+|Y|},
	\end{array}
	\right.
\end{align}
where N denotes the total number of samples, $y_i$, $\hat{y}_i$ respectively signify the true label and prediction. $|X|$ and $|Y|$ represent the ground truth and prediction, respectively. $\lambda_1$, $\lambda_2$ refer to the weights of loss functions, which are both set to 1 by default.
\section{Experiments}
\subsection{Datasets}
The dermoscopic images utilized for both training and evaluation were derived from the ISIC2018 Machine Learning Challenge (ISIC2018: Skin Lesion Analysis for Melanoma Detection). The training dataset consists of 8,010 samples, which are categorized into seven distinct disease classes. For the evaluation phase, a subset of 161 samples was selected from the original training set. The lesion images were sourced from the HAM10000 dataset, which is publicly accessible via both the archive gallery and standardized API endpoints. Throughout the ISIC2018 challenge, image data, along with corresponding diagnostic information and ground-truth labels, were made available for download. The competition is organized around three separate tasks, each addressing different facets of skin lesion analysis, providing a comprehensive evaluation of model performance in melanoma detection.
\\
\hspace*{1em}
The ISIC 2017 dataset is a well-known benchmark dataset for skin cancer analysis, consisting of a total of 2,750 dermoscopic images of skin lesions. These images are systematically divided into three subsets: 2,000 images allocated for training, 150 images designated for testing, and 600 images reserved for validation. One of the distinguishing features of this dataset is the significant variation in image resolutions, ranging from 540×722×3 to as large as 4499×6748×3. The high resolution of the images provides a substantial advantage over other dermoscopic image datasets, as it enables more precise lesion visualization, facilitates detailed feature extraction, and enhances the accuracy of segmentation and classification tasks. This makes the ISIC 2017 dataset a valuable resource for developing and evaluating advanced machine learning models in dermatological image analysis.

\subsection{Implementation Details}
The images from the ISIC2017 and ISIC2018 datasets are preprocessed to a resolution of 256×256 pixels before being fed into the model. Training is conducted with a batch size of 8, utilizing the AdamW optimizer \cite{81} with an initial learning rate of $1\times10^{-3}$. In order to facilitate dynamic learning rate adjustments throughout the training process, a Cosine-AnnealingLR scheduler \cite{82} is employed. All experiments are conducted on a single NVIDIA TESLA A100-PCIE-40GB GPU for 150 epochs, providing the computational power necessary for efficient training and evaluation.
\subsection{Results}
\textbf{Segmentation performamnce.} In order to demonstrate the effective of our proposed approach, we conducted a comparative analysis of xLSTM-VMUNet against other state-of-the-art methodologies. Specifically, they include R2U-Net \cite{83}, Attention R2U-Net \cite{84}, Attention-U-Net \cite{84}, U-Net \cite{16}, U-Net++ \cite{17}, Swin-U-Net \cite{85}, Res-U-Net \cite{86}, CaraNet \cite{87}, FANet \cite{88}, PraNet \cite{89}, VM-UNetv2 \cite{90}, UltraLight VM-Unet \cite{60}, TransUNet \cite{91}, VM-UNet \cite{53}, SkinMamba \cite{61} and UNeXt \cite{21}.\\
\captionsetup[table]{singlelinecheck=false}
\begin{table}[H]
	\centering
	\setlength{\tabcolsep}{20pt}
	\renewcommand{\arraystretch}{1}
	\setlength{\belowcaptionskip}{2pt}
	\begin{tabular}{ccc}
		\hline
		\textbf{Model} & \textbf{DSC$\uparrow$} & \textbf{IoU$\uparrow$} \\
		\hline
		R2U-Net \cite{83} & 0.6780 & 0.5760 \\
		Attention R2U-Net \cite{84} & 0.7020 & 0.6128 \\
		Attention-U-Net \cite{84} & 0.7210 & 0.6231 \\
		U-Net+ \cite{16} & 0.7030 & 0.6159 \\
		U-Net++ \cite{17} & 0.7300 & 0.6340 \\
		Swin-U-Net \cite{85} & 0.7500 & 0.6500 \\
		Res-U-Net \cite{86} & 0.7520 & 0.6542 \\
		CaraNet \cite{87} & 0.7550 & 0.6603 \\
		FANet \cite{88} & 0.7800 & 0.6851 \\
		PraNet \cite{89} & 0.7950 & 0.7024 \\
		VM-UNetv2 \cite{90} & 0.8330 & 0.7340 \\
		UltraLight VM-UNet \cite{60} & 0.8500 & 0.7511 \\
		TransUNet \cite{91} & 0.8680 & 0.7800 \\
		VM-UNet \cite{53} & 0.8800 & 0.7832 \\
		SkinMamba \cite{61} & 0.9000 & 0.8077 \\
		UNeXt \cite{21} & 0.9080 & 0.8290 \\
		xLSTM-VMUNet & \textbf{0.9200} & \textbf{0.8525} \\
		\hline
	\end{tabular}
	\caption{\centering Comparison with state-of-the-art methods on the ISIC2017 dataset, \\5-fold cross-validation. (Bold indicates the best.)}
	\label{Tab.1}
\end{table}

\begin{table}[H]
	\centering
	\setlength{\tabcolsep}{20pt}
	\renewcommand{\arraystretch}{1}
	\setlength{\belowcaptionskip}{2pt}
	\begin{tabular}{ccc}
		\hline
		Model & DSC$\uparrow$ & IoU$\uparrow$ \\
		\hline
		R2U-Net \cite{83} & 0.6790 & 0.5810 \\
		Attention R2U-Net \cite{84} & 0.7260 & 0.5920 \\
		Attention-U-Net \cite{84} & 0.8205 & 0.7346 \\
		U-Net+ \cite{16} & 0.8403 & 0.7455 \\
		U-Net++ \cite{17} & 0.8496 & 0.7512 \\
		Swin-U-Net \cite{85} & 0.8523 & 0.7528 \\
		Res-U-Net \cite{86} & 0.8560 & 0.7562 \\
		CaraNet \cite{87} & 0.8702 & 0.7822 \\
		FANet \cite{88} & 0.8731 & 0.8023 \\
		PraNet \cite{89} & 0.8754 & 0.7874 \\
		VM-UNetv2 \cite{90} & 0.8796 & 0.7851 \\
		UltraLight VM-UNet \cite{60} & 0.8820 & 0.7890 \\
		TransUNet \cite{91} & 0.8891 & 0.8051 \\
		VM-UNet \cite{53} & 0.8975 & 0.8142 \\
		SkinMamba \cite{61} & 0.8976 & 0.8143 \\
		UNeXt \cite{21} & 0.9030 & 0.8261 \\
		xLSTM-VMUNet & \textbf{0.9100} & \textbf{0.8349} \\
		\hline
	\end{tabular}
	\caption{\centering Comparison with state-of-the-art methods on the ISIC2018 dataset, \\5-fold cross-validation. (Bold indicates the best.)}
	\label{Tab.2}
\end{table}
\newcommand{\Tabref}[1]{Tab.\ref{#1}}
\captionsetup[figure]{singlelinecheck=false}
\begin{figure}[H]
	\centering
l	\includegraphics[width=1\linewidth]{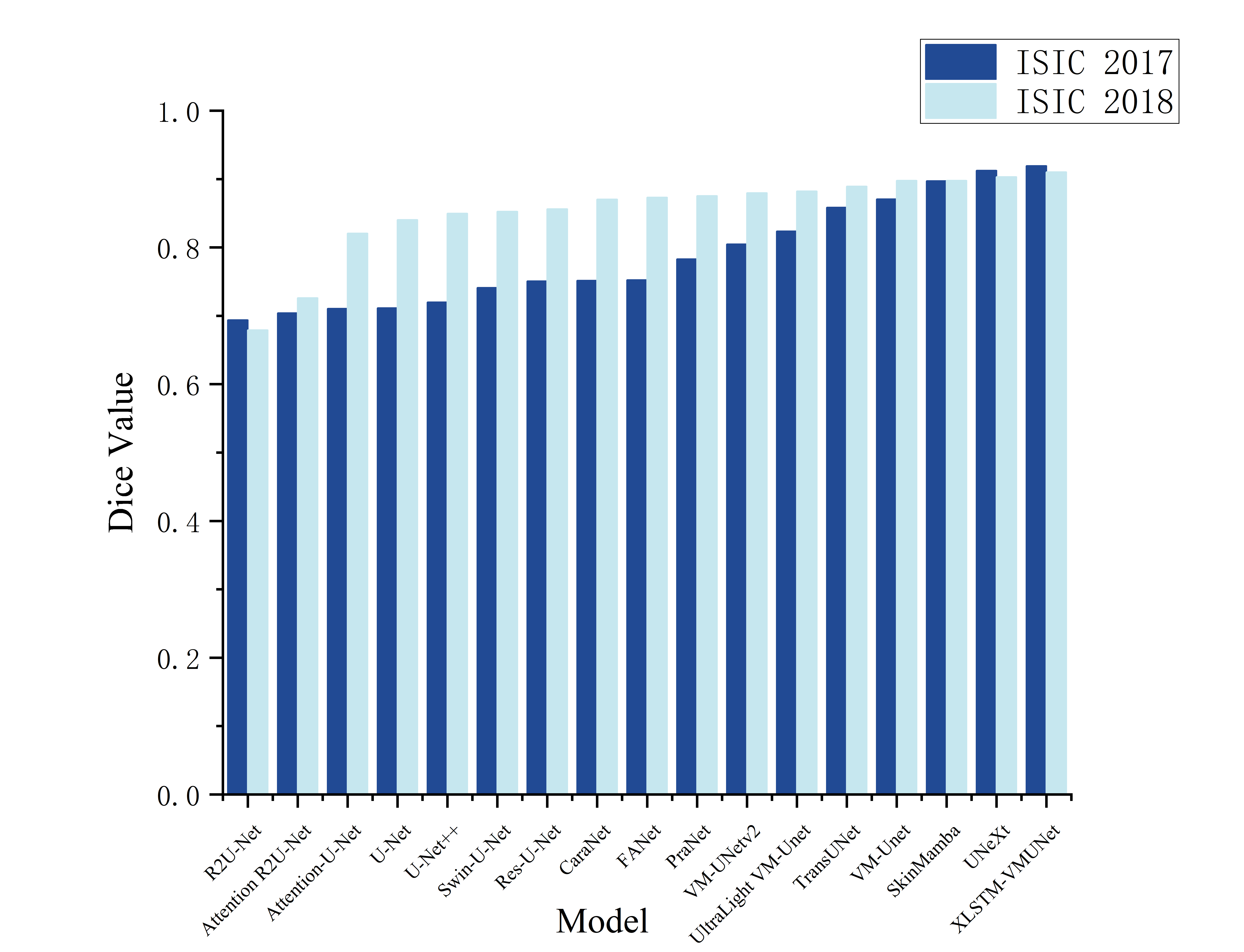}
	\caption{\centering Comparison of Dice values across different models \\on the ISIC2017 and ISIC2018 datasets.}\label{Fig.3}
\end{figure}
\Tabref{Tab.1} and \Tabref{Tab.2} provide a detailed comparative analysis of the segmentation performance on both the ISIC 2017 and ISIC 2018 datasets. The results clearly demonstrate that the proposed xLSTM-VMUNet achieves superior performance compared to other state-of-the-art models, particularly in terms of key evaluation metrics such as the Dice Similarity Coefficient (DSC) and Intersection over Union (IoU). The consistently higher DSC and IoU values indicate that our model effectively captures lesion boundaries and enhances segmentation accuracy. Furthermore, \Myref{Fig.3} offers a visual comparison of the segmentation outcomes based on the Dice metric, illustrating the qualitative advantages of our approach. The visualized results highlight the model’s ability to generate more precise and well-defined segmentation masks, further validating its effectiveness in skin lesion analysis.
\captionsetup[figure]{singlelinecheck=false}
\begin{figure}[H]
	\centering
	\includegraphics[width=0.8\linewidth]{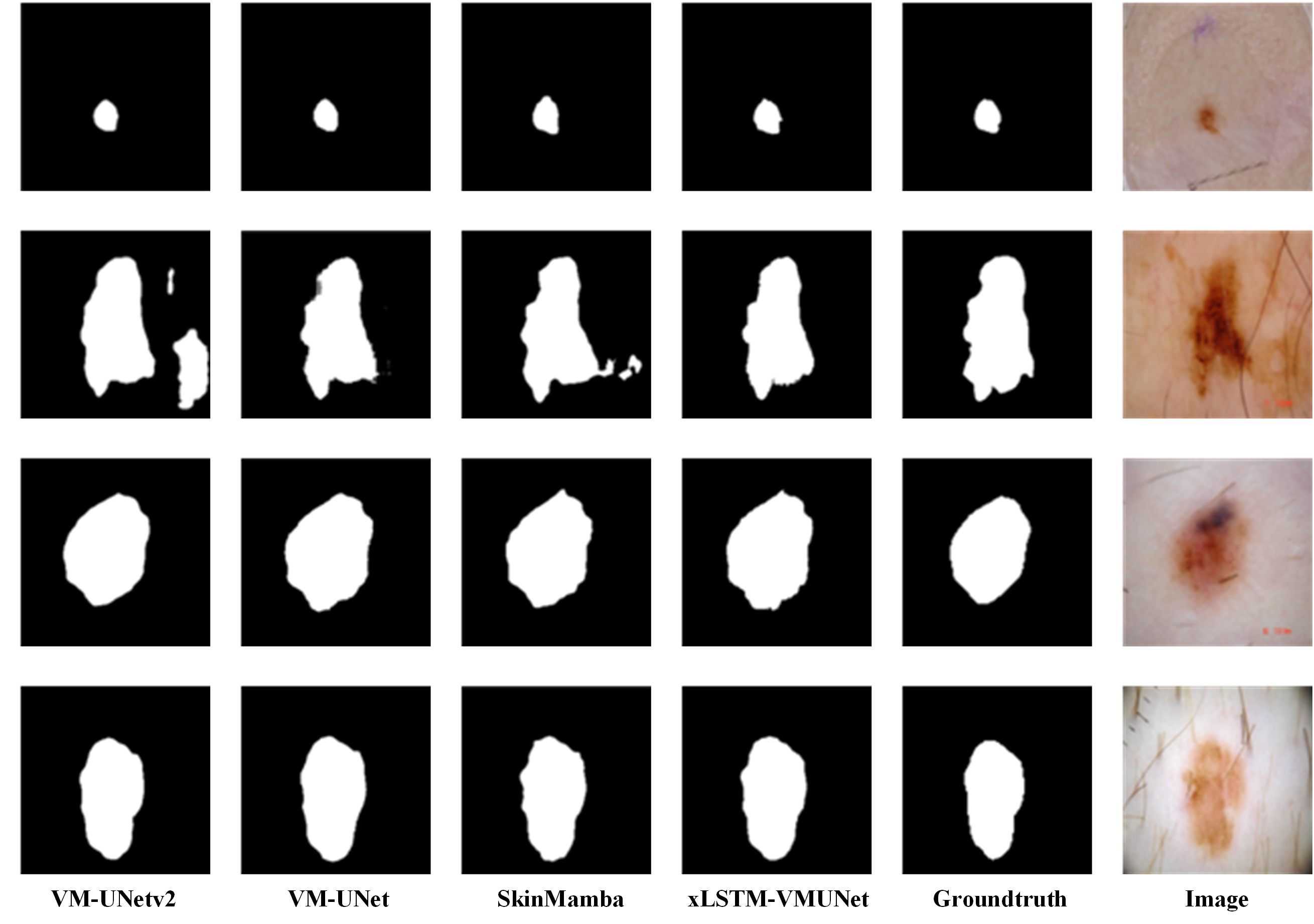}
	\caption{\centering Visualized segmentation results of the proposed xLSTM-VMUNet \\against other state-of-the-arts over the ISIC2018 dataset.}\label{Fig.5}
\end{figure}
\indent{As illustrated in \Myref{Fig.5}, the proposed xLSTM-VMUNet model demonstrates a significantly enhanced sensitivity to fine-grained lesion details when compared to conventional segmentation methods. This improvement can be attributed to its ability to effectively capture multi-scale contextual information while leveraging temporal dependencies, which enables the model to discern subtle variations in lesion characteristics with greater accuracy. By integrating these advanced mechanisms, xLSTM-VMUNet not only enhances the precision of lesion boundary delineation but also mitigates common segmentation challenges such as over-segmentation and under-segmentation. This capability is particularly crucial in the field of medical image analysis, where precise lesion segmentation plays a fundamental role in ensuring reliable clinical assessments, aiding early diagnosis, and optimizing treatment planning. The model’s superior performance in identifying intricate lesion structures underscores its potential for advancing automated medical image interpretation and improving diagnostic workflows in dermatology and other related medical domains. The results of the ablation experiments are shown in \Tabref{Tab.3} and \Tabref{Tab.4}.}\
\begin{table}[H]
	\centering
	\setlength{\tabcolsep}{20pt}
	\renewcommand{\arraystretch}{1}
	\setlength{\belowcaptionskip}{2pt}
	\begin{tabular}{cccccc}
		\hline
		Ver. & sLSTM & mLSTM & DSC$\uparrow$ & IoU$\uparrow$ \\
		\hline
		Ver 1 & $\times$ & $\times$ & 0.8800 & 0.7832 \\
		Ver 2 & $\checkmark$ & $\times$ & 0.9082 & 0.8215 \\
		Ver 3 & $\times$ & $\checkmark$ & 0.9087 & 0.8219 \\
		Ver 4 & $\checkmark$ & $\checkmark$ & \textbf{0.9200} & \textbf{0.8525} \\
		\hline
	\end{tabular}
	\caption{\centering Ablation experiments on the ISIC2017 dataset: impact of individual contributions on segmentation performance of xLSTM-VMUNet. (Bold indicates the best.)}
	\label{Tab.3}
\end{table}
\begin{table}[H]
	\centering
	\setlength{\tabcolsep}{20pt}
	\renewcommand{\arraystretch}{1}
	\setlength{\belowcaptionskip}{2pt}
	\begin{tabular}{cccccc}
		\hline
		Ver. & sLSTM & mLSTM & DSC$\uparrow$ & IoU$\uparrow$ \\
		\hline
		Ver 1 & $\times$ & $\times$ & 0.8975 & 0.8142 \\
		Ver 2 & $\checkmark$ & $\times$ & 0.9013 & 0.8238 \\
		Ver 3 & $\times$ & $\checkmark$ & 0.9098 & 0.8345 \\
		Ver 4 & $\checkmark$ & $\checkmark$ & \textbf{0.9100} & \textbf{0.8349} \\
		\hline
	\end{tabular}
	\caption{\centering Ablation experiments on the ISIC2018 dataset: impact of individual contributions on segmentation performance of xLSTM-VMUNet. (Bold indicates the best.)}
	\label{Tab.4}
\end{table}
\section{Conclusion}
In this study, we propose the xLSTM-VMUNet model and demonstrate the benefits of effectively combining Mamba with xLSTM: xLSTM-VMUNet can not only focus on extracting spatial features from images, especially the structure and characteristics of skin lesions, but also enhance contextual understanding of the model, allowing it to better handle complex structures in medical images. This dual integration significantly improves the accuracy of skin lesion segmentation and enhances the computational efficiency of the model. Experimental results indicate that on the ISIC2017 dataset, xLSTM-VMUNet outperforms UNeXt with a 1.20\% improvement in the DSC metric and a 2.35\% improvement in the IoU metric, on the ISIC2018 dataset, xLSTM-VMUNet outperforms UNeXt with a 0.70\% improvement in the DSC metric and a 0.88\% improvement in the IoU metric. Compared to the baseline model, VM-UNet, xLSTM-VMUNet shows a 4.00\% increase on DSC and 6.93\% increase on IoU on the ISIC2017 dataset, 1.25\% increase on DSC and 2.07\% increase on IoU on the ISIC2018 dataset, with faster convergence and consistently high segmentation performance.

\end{document}